\documentclass[aps,prd,twocolumn,superscriptaddress,amsfont,amssymb,amsmath,nofootinbib,showpacs,balancelastpage]{revtex4-1}

\usepackage{graphicx,longtable,natbib,mathrsfs,color}
\usepackage{txfonts}

\newcommand{\bs}{\boldsymbol}
\newcommand{\diff}{{\mathrm d}}
\newcommand{\Cov}{\mathsf{C}}
\newcommand{\Fish}{\mathsf{F}}

\newcommand{\R}{\mathcal{R}}

\begin{document}

\addtolength{\hoffset}{-0.525cm}
\addtolength{\textwidth}{1.05cm}
\title{Nonparametric reconstruction of dynamical dark energy via observational Hubble parameter data}

\author{Hao-Ran~Yu} \affiliation{Department of Astronomy, Beijing Normal University, Beijing 100875, China}
\author{Shuo~Yuan} \affiliation{Department of Astronomy, Beijing Normal University, Beijing 100875, China}
\author{Tong-Jie~Zhang}
\affiliation{Department of Astronomy, Beijing Normal University, Beijing 100875, China}


\begin{abstract}
We study the power of current and future observational Hubble parameter data (OHD) on non-parametric estimations of the dark energy equation of state, $w(z)$. We propose a new method by conjunction of principal component analysis (PCA) and the criterion of goodness of fit (GoF) criterion to reconstruct $w(z)$, ensuring the sensitivity and reliability of the extraction of features in the EoS. We also give an new error model to simulate future OHD data, to forecast the power of future OHD on the EoS reconstruction. The result shows that current OHD, despite in less quantity, give not only a similar power of reconstruction of dark energy compared to the result given by type Ia supernovae, but also extend the constraint on $w(z)$ up to redshift $z\simeq2$. Additionally, a reasonable forecast of future data in more quantity and better quality greatly enhances the reconstruction of dark energy.
\end{abstract}


\maketitle

\section{Introduction}\label{sec.intro}
The accelerating expanding universe is explained by the dark energy \citep{1998AJ....116.1009R,1999ApJ...517..565P} and one of the main challenges of the modern physical cosmology is to study the nature of dark energy \citep{2004PhRvL..92x1302W}. Its evolution and dynamical properties are characterized by its equation of state (EoS) $w\equiv p/\rho$, the ratio of its pressure and energy density. A cosmological constant is usually the simplest model to explain dark energy and its EoS does not depend on redshift $z$, and remains constant $w(z)=-1$. Variations of the dark energy models, such as Quintessence, phantom, Quintoms etc., are also used to explain the dark energy. Studying dark energy in a parametrized way, such as CPL (Chevallier-Polarski-Linder) parametrization \citep{2001IJMPD..10..213C,2003PhRvL..90i1301L}, may induce misleading results due to our prior assumptions of function forms of EoS, then we may do null-test diagnostics on dark energy \citep{2008PhRvD..78j3502S} or use model-independent analyses to reconstruct EoS non-parametrically \citep{2012PhRvL.109q1301Z,2004ApJ...612..652D,2008ApJ...677....1D}. This has been achieved by astronomical data, such as luminosity distances $d_L$ of type Ia supernovae (Ia SN) \citep{2003PhRvL..90c1301H,2010PhRvL.104u1301C}. Although we have plenty of Ia SN data, it requires the second derivative of $d_L$ respect to redshift $z$, and a complicated form of reconstruction equation. While, observational Hubble parameter data (OHD) $H(z)$ reconstruct $w(z)$ via a simpler way (Eq.(\ref{eq.wz})) and require only the first derivative of $H(z)$, so it is more stable and less sensitive to the error of the data. From the last decade, OHD are proved to be very powerful in the constraint of cosmological parameters \citep{2007MPLA...22...41Y}, it responses the cosmic expansion history directly and can be measured by various methods, e.g. cosmic chronometers \citep{2002ApJ...573...37J}, baryonic acoustic oscillation (BAO) peaks \citep{2009MNRAS.399.1663G}, and even other proposed Sandage-Loeb (SL) probes \citep{1962ApJ...136..319S,1998ApJ...499L.111L} and standard siren by gravitational waves. On the way of obtaining more data from observations, to simulate sets of future OHD appropriately is also instructive at the current stage in exploring the quality of EoS reconstruction in the near future. To do this, we need an error model giving the redshift distribution, the offsets from the theoretical value, and the sizes of their error bars, such as in \cite{2011ApJ...730...74M}. Here we construct a new and more accurate error model for next generation surveys, carefully estimating potential power of observation in different methods.

For either observational or our simulated data, nonparametrically reconstructing $H(z)$ and $w(z)$ from data confronts the trading between extracting more features and avoiding the over-fitting of errors -- if we use less parameters, we are losing details of the underlying model, while using more parameters increases the detectability of such details, but it brings the risk of being polluted by the errors. Principle component analysis (PCA) is usually used to exclude error-induced oscillations in the reconstructing processes \citep{2003PhRvL..90c1301H}, and information criteria \citep{2007MNRAS.377L..74L} are often used for choosing models \citep{2010PhRvL.104u1301C}. We also find that the goodness of fit (GoF) is a effective probe of over-fitting. The conjunction of PCA and GoF criterion works naturally to deal with this tradeoff.

The rest of the paper is organized as follows. We present our reconstruction method in Sec.\ref{sec.method}, and briefly illustrate the current OHD and forecast future OHD in Sec.\ref{sec.data}. We show the results by our method in Sec.\ref{sec.results} and conclude in Sec.\ref{sec.discussion}.

\section{Method}\label{sec.method}
We briefly show the mathematics of the reconstruction method in Sec.\ref{sec.intro}. To test the method we use the new error model to simulated data, which is shown in Sec.\ref{sec.errormodel}, and the GoF criterion is discussed in Sec.\ref{sec.gof}.
\subsection{Reconstruction}\label{sec.reco}
In general relativity, the expansion of the universe is affected by the densities of the components and the state of dark energy:
\begin{equation}
	H(z)^2=H_0^2[ \Omega_{\rm M}(1+z)^3+\Omega_{\rm K}(1+z)^2+\Omega_\Lambda \mathfrak{g}(z)],
\end{equation}
where $H_0$, $\Omega_{\rm M}$, $\Omega_{\rm K}$, $\Omega_\Lambda$ are the Hubble constant, the current energy density in pressureless non-relativistic matter, curvature and a smooth dark energy component: $\Omega_{\rm M}+\Omega_{\rm K}+\Omega_\Lambda=1$, and
	$\mathfrak{g}(z)=\exp\left[3\int_0^z (1+w(z'))/(1+z')\diff z'\right]$.
Here we neglect the current energy density of radiation $\Omega_{\rm R}$ and assume the universe to be dominated by non-relativistic matter and dark energy. Expressing $w(z)$ in terms of $H(z)$ and $H'(z)\equiv \diff H/\diff z$ and we get the reconstruction equation:
\begin{equation}\label{eq.wz}
	w(z)=\dfrac{3H^2-2(1+z)HH'-\Omega_{\rm K}H_0^2(1+z)^2}	{3H_0^2(1+z)^2[\Omega_{\rm M}(1+z)+\Omega_{\rm K}]-3H^2}.
\end{equation}
Given a set of $H(z)$ data $\bs{y}$, we fit them by a smooth analytical function then use it and its first derivative to reconstruct $w(z)$ by Eq.(\ref{eq.wz}). Assume the data $\bs{y}=(y_1,y_2,...,y_n)^{\mathsf T}$ with covariance matrix $\Cov$. By choosing a set of $N$ ($N<n$) primary basis functions such as polynomials, rational functions or wavelets: $\bs{X}=(\bs{x}_1^{\mathsf T},\bs{x}_2^{\mathsf T},...,\bs{x}_N^{\mathsf T})$, where $\bs{x}_i$ is a row vector representing the $i$th basis, we fit $\bs{y}$ by linear least squares. Simply minimize the weighted squared residual $\R=(\bs{y}-\bs{X\beta})^{\mathsf T}\Cov^{-1}(\bs{y}-\bs{X\beta})$ by solving $\partial\R/\partial\bs\beta=\bs{0}$, and we get the coefficient vector
\begin{equation}\label{eq.beta}
	\bs\beta=(\bs{X}^{\mathsf T}\Cov^{-1}\bs{X})^{-1}\bs{X}^{\mathsf T}\Cov^{-1}\bs{y}.
\end{equation}

By enough realizations from the error model, we diagonalize the inverse covariance matrix of $\bs\beta$ by finding its $N$ eigenvalue and eigenvectors:
$\Fish={\mathrm{Cov}}^{-1}(\bs{\beta})={\mathsf {E\Lambda E}}^{\mathsf{T}}$,
where ${\mathsf E}=(\bs{e}_1^{\mathsf T},\bs{e}_2^{\mathsf T},...,\bs{e}_N^{\mathsf T})$
rearranges the $N$ primary basis functions to $N$ new orthogonal eigenbasis
$\bs{U}=\bs{X}{\mathsf E}$
with the corresponding eigenvalues
${\mathsf{diag}}(\mathsf\Lambda)=(\lambda_1,\lambda_2,...,\lambda_N)^{\mathsf T}$.
The first $M$ ($M\leq N$) eigenmodes
$\bs{U}_M=\bs{X}{\mathsf E}_M$, where
${\mathsf E}_M=(\bs{e}_1^{\mathsf T},\bs{e}_2^{\mathsf T},...,\bs{e}_M^{\mathsf T})$, the principal components of the data,
are usually of our interest, reflecting the main features of data. Conversely, the rest components are induced by the random offsets in the data and may induce overfittings. After the rearrangement, we solve Eq.(\ref{eq.beta}) again and get new coefficients for the new basis:
\begin{equation}\label{eq.beta_M}
	\bs\beta_M=(\bs{U}_M^{\mathsf T}\Cov^{-1}\bs{U}_M)^{-1}\bs{U}_M^{\mathsf T}\Cov^{-1}\bs{y}.
\end{equation}
Because we have enough realizations to estimate $\rm{Cov}(\bs{\beta})$, and any single new realization or observational data, having similar configurations, their true features can still be extracted and the nuisance features are excluded. With the reliable smooth $H(z)$, getting $w(z)$ is straightforward from Eq.(\ref{eq.wz}) and its errors can be estimated from realizations.

\subsection{An optimal error model}\label{sec.errormodel}
Usually simulated offsets and errors are regarded as Gaussian distributed \citep{2011ApJ...730...74M}. More generally, the error, a random variable $\sigma$, is determined by several factors $\bs{t}=(t_1,t_2,...,t_k)^{\mathsf T}$, where these factors are assumed to be Gaussian distributed with covariance $\bs\Sigma_t$. These factors are different sources of errors that jointly contribute to $\sigma$ by a quadratic form $\sigma^2=\bs{t}^{\mathsf T}\mathsf{D}\bs{t}$, and the matrix $\mathsf{D}$ denotes the relation between factor and the final error, up to the second order accuracy. In such case, $\sigma$ has a Nakagami $m$-distribution $f_m(x;m,\Omega)$ (a scaling transform of generalized-$\chi$-distribution with parameters $\bs\Sigma_t$ and $\mathsf{D}$) \citep{1977JRSS...26.92H}, and has a complicated form. We can never estimate the full contribution of these factors (weather, telescope, device, recording during observation, as long as data reduction and systematic errors etc.), however, we simplify the problem by assuming $t_i\sim\mathcal{N}(0,1)$, $\bs\Sigma_t=\mathsf{D}=\bs{I}$: they are independent, having same importance, and contribute to $\sigma^2$ additively. In this case, $\sigma\sim\chi_k$. More realistically, there are few dominant factors, of much greater importance ($\bs\Sigma_t\neq \bs{I}$), jointly contribute to the final error. This effectively cause a reduction in the degree of freedom, $k\rightarrow k'$ ($k'<k$) and a scaling transform ($\chi_{k'}\rightarrow A\chi_{k'}$). Practically, neglecting the mathematics details, we find that the $m$-distribution $f_m(x;m,\Omega)$ ($m=k'/2$, $\Omega=A^2/k'$) well matches the relative error $\sigma_H/H$'s distribution and does not depend on redshift $z$.

The redshift distribution is generated according to the configuration of each method of measurement. For large enough sample, we find that uniform distributed or evenly spaced samples are good approximations. For errors and offsets, we assume the data points are independent, so the relative error is successively generated from the correspond distribution $\sigma/H_\star(z)\sim f_i$ where $f_i$ is the relative error distribution for the $i$th kind measurement. This means that, we are ready for a measurement, without knowing the true value $y_\star=H_\star(z)$, but the quality (error bar) is predetermined. For $\langle f_i\rangle\ll 1$, the resulting measurement should have the distribution $y\sim\mathcal{N}(y_\star,\sigma)$, and it gives the simulated data.

\subsection{Goodness of fit}\label{sec.gof}
The remaining problem is how to choose the number of primary basis functions $N$ and the number of reserved principal components $M$ appropriately. The choice of $(N,M)$ directly relates to the upper limit of the ability of how complex we can detect the features of the underlying model. If we suppose that the fitted model is just the underlying model, the offset $\bs{y}-\bs{y}_\star$, having the distribution of $\mathcal{N}(0,\sigma)$, contributes to the residual $R_\star$ by $(\mathcal{N}(0,1))^2$ for each independent data point, and $\R_\star$ has $\chi^2$-distribution $\R_\star\sim\chi^2_n$. Thus
\begin{equation}
	\langle {\rm GoF}_\star \rangle\equiv\left\langle \dfrac{\R_\star}{n} \right\rangle=\left\langle \dfrac{(\bs{y}-\bs{y}_\star)^{\mathsf{T}}\Cov^{-1}(\bs{y}-\bs{y}_\star)}{n} \right\rangle=1.
\end{equation}
Here we use $\star$ to denote that the fitted model is replaced by the underlying model, and also define $R_\star/n$ as the ``goodness of `fit'" for underlying model ${\rm GoF}_\star$. Going back to the resulting fitted model, define the GoF in our circumstance as\footnote{Note that, although it is usually to scale $R_M$ by the degree of freedom $\nu=n-M-1$, here we instead use $n-1$ to scale $R_M$ for estimated underlying model from $n$ data samples. Because various choice of primary basis functions and inertial complexity of underlying model lead to different numbers of parameters needed. Even, given a fixed underlying model, resulting $N$ and $M$ depend on $\bs{X}$. Thus the expression of GoF should not be scaled by anything in terms of $N$ nor $M$ in this circumstance.}
\begin{equation}
	{\rm GoF}\equiv\dfrac{\R_M}{n-1},
\end{equation}
where $\R_M$ is the least square $\R_{\rm min}$ given by the first $M$ principal components $\bs{U}_M$:
\begin{equation}\label{eq.chi2_M}
	\R_M=(\bs{y}-\bs{U}_M\bs\beta_M)^{\mathsf T}\Cov^{-1}(\bs{y}-\bs{U}_M\bs\beta_M).
\end{equation}

Statistically $\langle {\rm GoF}\rangle$ should also be unity. If it is less than one, it indicates the overfitting of offsets due to the measurement error; while if GoF is larger than one for a great amount, it means that we have not yet capture the full features of data, i.e. the model is too simple to fit all the features. We use GoF as the indicator to determine the complexity of our fitted model. Here, the complexity of the fitted model is less or equal than the true underlying model, because errors and offsets lower the detectability of the underlying model. More specifically, for a very complex oscillating $w(z)$ as a example, it always let its generated data to have ${\rm GoF}_\star\simeq 1$, and we are supposed to use higher $N$ and $M$ to fit its complex features. However, if the generated data are dominated by noise, such configuration ($N$ and $M$) leads to illness in fitting low quality data, because the errors and offsets are more dominant than the true features and thus be amplified and any subtler features are covered. As a result, the reconstructed model is no longer the underlying model and usually has much smaller GoF (${\rm GoF}<{\rm GoF}_\star\simeq 1$).

For given large enough $N$, fully usage of all eigenmodes ($M=N$) leads to the overfitting, and we degrade the complexity of model by reduce $M$ consecutively while examine $\langle {\rm GoF}\rangle$ over realizations, until we find the last $M$ to have $\langle {\rm GoF}\rangle>1$ --  it is still save to use this grade of complexity. Insufficient number of primary basis functions $N$ causes two consecutive $M$'s to have a great difference in $\langle {\rm GoF}\rangle$ -- skipping the range of appropriate fitting ($\langle {\rm GoF}\rangle\simeq 1$), as the eigenmodes are poorly determined from limited number of primary functions, or the primary functions $\bs{X}$ are chosen improperly. In these cases we should either increase $N$ or choose more suitable $\bs{X}$.

Without a clear knowledge of the essence of dark energy, the underlying complexity of $w(z)$ results in enumerable forms $H(z)$'s, so generally there is not an optimal choice of primary function basis $\bs{X}$. However, we can still choose some popular, reasonable models to see how different $\bs{X}$'s have effect on the fitting. Usually polynomials are more effective than other rational functions for not-too-complex models, and they form stable eigenmodes that are invariant for different values of $N$ and various underlying $H(z)$, even if the underlying $H(z)$ is very oscillatory.

\begin{figure}
\centering
\includegraphics[width=0.5\textwidth]{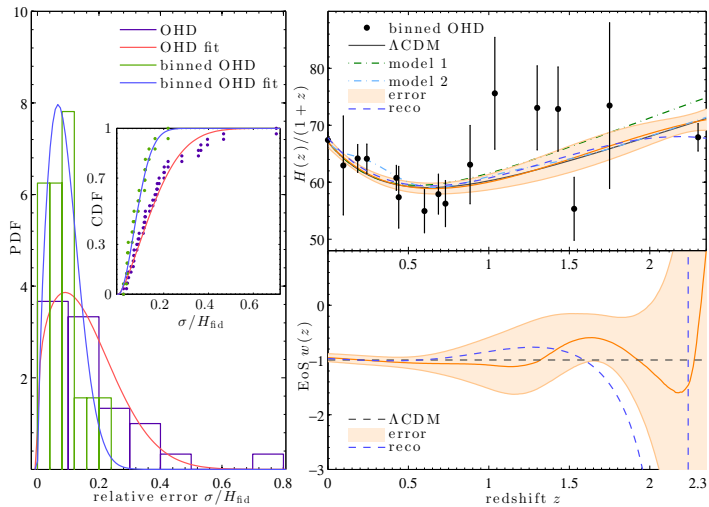}
\caption{(\textit{left}:) Relative error distribution for OHD. The purple and green histograms show the relative errors of unbinned 28 OHD and binned 15 OHD, and the red and blue curves are their best fits by a $m$-distribution. Their counterpart CDF and fitted CDF are shown in the inset. (\textit{right}:) Binned OHD and their reconstructions on $H(z)$ (\textit{top}) and $w(z)$ (\textit{bottom}). The shaded orange areas show the $1\sigma$ error regions if we assume a $\Lambda$CDM model. Dashed blue lines are the results from binned OHD.}
\label{fig:1}
\end{figure}

\section{Available and future Hubble parameter data}\label{sec.data}
Current OHD are obtained primarily by the method of cosmic chronometer
\citep{2005PhRvD..71l3001S,2010ApJS..188..280S,2012JCAP...08..006M,2012arXiv1207.4541Z}. Other methods to extract $H(z)$ are by the observations of BAO peaks\citep{2009MNRAS.399.1663G,2012MNRAS.425..405B}, Ly$-\alpha$ forest of luminous red galaxies (LRGs)  \citep{2013A&A...552A..96B}. The last of which extended the current OHD deep to $z=2.3$.

For the real data applied to our method, it is straightforward to test their ${\rm GoF}_\star$. Assuming the independency of 28 measurements, to calculate the residual between the data and a fiducial theoretical $\Lambda$CDM (cold dark matter with a constant cosmological constant $\Lambda$) model gives ${\rm GoF}_\star=0.62<1$, meaning that for independent data with current error level, their offsets would have been larger. It indicates that the independency assumption for the current 28 OHD is not proper. However, we do not have a good estimation of the off-diagonal elements of the OHD's covariance matrix. To weigh the power of current OHD and compare with simulations, we rebin the data from LRGs that are with close redshifts, and get finally 15 measurements of OHD with smaller error bars (see figure 1). Now for the binned data ${\rm GoF}_\star=1.02$ which is close to independent measurements. We have larger residual contributed to $\R$ statistically from each binned datum and it is reasonably from the error. Although this is only a rough estimation, it is still more accurate than the 28 data points with no knowledge of their covariance. The binned and unbinned error distributions for OHD from well match the shape of $m$-distribution (see the left panel of Fig.\ref{fig:1}). We also assume the relative error of future data to have such distribution. 

There are several ways to enhance the quality of OHD, and of which deeper-redshift, more-complete-sky-coverage LRG survey and spectroscopic observations of those identified LRGs give remarkable improvement on the two methods on LRG. For example, the 2SLAQ\footnote{\url{http://www.2slaq.info/}} has provided a LRG catalogue with the redshift range from 0.3 to 0.9 with 180 $\deg^{2}$ coverage of the sky. Considering a future LRG survey with more than half sky coverage with redshift range $z\sim 2.3$, it may give several millions of LRGs and effectively enlarges the number of OHD data points or lowers their errors. By an LRG sample binning strategy \citep{2005PhRvD..71l3001S}, up to 100 OHD measurements can be extracted with $20\%$ of present error level.

\begin{figure}
\centering
\includegraphics[width=0.5\textwidth]{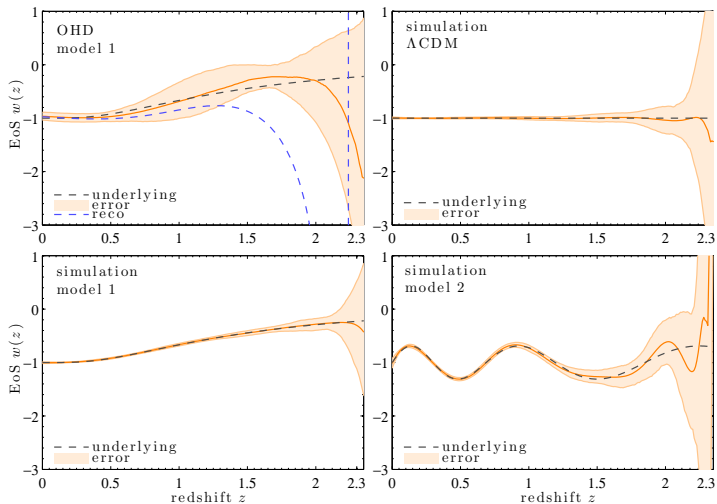}
\caption{(\textit{top-left}:) Reconstruction of $w(z)$ by OHD's redshift and error configurations, but assuming model-1. Also shown in blue dashed line is the reconstruction of OHD set, which is obviously inconsistent with model-1. The rest three panels show the reconstruction of $w(z)$ from simulated future data, with the underlying model being $\Lambda$CDM, model-1 and model-2.}
\label{fig:2}
\end{figure}

Other methods are also potential: \cite{2012MNRAS.425..405B} give three OHD by processing the WiggleZ\footnote{\url{http://wigglez.swin.edu.au/site/}} survey, fitting BAO peak parameter and the 2D power spectrum at three redshift slices. The raw WiggleZ samples are 158741 galaxies in the range $z=(0.2,1.0)$ with 800 $\deg^2$ coverage of the sky. We can also forecast a more-than-half-sky survey like WiggleZ, which offers up to five million targets and thus lowers the error level to several percent of present level. The more ambitious experiment for Sandage-Loeb signal(SLS), proposed by \cite{1962ApJ...136..319S,1998ApJ...499L.111L}, may further deepen the redshift range of OHD. It measures the quasar Ly$-\alpha$ forest in a separation of few decades by the extremely large telescope like the European Extremely Large Telescope (E-ELT)\footnote{\url{http://www.eso.org/public/teles-instr/e-elt.html}}. The upcoming CODEX (COsmic Dynamics and EXo-earth experiment)\footnote{\url{http://www.iac.es/proyecto/codex/}} is based on the E-ELT and offers a measurement of SL signal. If succeed, it can help us to explore the redshift from 2 to 5 covering the ``redshift dersert" and give useful data for the expanding history of the universe \citep{2007PhRvD..75f2001C}. The CODEX group provided a full design of observing the SL signal and the prediction of the statistical error of SLS \citep{2008MNRAS.386.1192L}. We use the SL signal, with an error estimation by \cite{2012PhRvD..86l3001M} (15 years observational interval is assumed), as an optional simulated data to study its impact on the result.

\section{Results}\label{sec.results}
We set cosmological parameters $\Omega_{\rm M}$, $\Omega_{\rm K}$, $\Omega_\Lambda$ and $H_0$ as in the lasted Planck data release \citep{2013arXiv1303.5076P}. For the diagnostic EoS models, we choose $\Lambda$CDM with $w_\Lambda(z)=-1$ and other two arbitrary models $w_1(z)=-1/2+{\rm erf}({\ln(2z/e)})$ and $w_2(z)=-1-0.31\sin(12\ln(1/(1+z)))$, where in model-\textsf{1} $w_1$ smoothly variates from -1 to 0 as $z$ increases, while in model-\textsf{2} $w_2$ is very oscillatory. The theoretical $H(z)$ curves based on these two models are shown in dash-dotted lines in the top-right panel of Fig.\ref{fig:1}. We also show the dashed blue line, representing the best-fit analytical $H(z)$ curve by using 15 binned OHD. Its reconstruction of $w(z)$ is shown in the bottom-right panel of Fig.\ref{fig:1}.

We do not show errors for these two dashed blue lines -- usually Monte-Carlo realizations are run on each data point: $y_{\rm MC}\sim\mathcal{N}(y,\sigma)$ to get the statistical properties. However, recall that $y\sim\mathcal{N}(y_\star,\sigma)$, so $y_{\rm MC}\sim\mathcal{N}(y_\star,\sqrt{2}\sigma)$: its error is amplified, and also we have only one realization -- real observation, $y_{\rm MC}$ is biased by $y-y_\star$, due to the cosmic variance. By such reason we can only get an error estimation based on a supposed underlying model, e.g. $\Lambda$CDM -- assuming $y$ is just one realization of $y_\star$, and we calculate the statistics from other realizations from $y_\star$: $y_{\rm MC}\sim\mathcal{N}(y_\star,\sigma)$, and see if the reconstruction from the real data is within the error region of $y_{\rm MC}$. We use our error model to simulate data and do this Monte-Carlo realization. Here we use only cosmic chronometer data, with 100 independent measurements, 20\% of present error level, evenly distributed on $0<z<2.3$. The results are shown with expectations (orange lines) and $1\sigma$ their error regions (translucent orange areas) in Fig.\ref{fig:1}. For $H(z)$, it reasonably covers the $H(z)$ by underlying $\Lambda$CDM. $w(z)$ is confident when $z\lesssim 1.5$, while beyond this range the reconstruction is biased due to the scarceness of data, and when $z\simeq 2$ it is even hopeless because $H'$ is poorly determined. Comparing the the result (blue dashed line) with the error region, we still see a obvious $\sim1.5\sigma$ deviation from $\Lambda$CDM at $1.6\lesssim z\lesssim 2$ and it does not correlate with the bias of the central line of the error region. Although we note that this feature is also familiar to the result by Ia SNe \citep{2010PhRvL.104u1301C}, it could not be verified by few data with low quality around $1<z<2$ and a single datum (better quality) at $z=2.3$. Here a single datum at $z=2.3$ \citep{2013A&A...552A..96B}, which is by BAO features in the redshift range $2.1<z<3.5$, plays an important role in determining $H$ and $H'$ which sensitively tune the resulting $w(z)$. Thus further more precise measurements at high redshift are needed. Our forecast of future data is proved to be able to verify or refute this deviation. The top-left panel of Fig.\ref{fig:2} shows another Monte-Carlo simulation, with $\Lambda$CDM replaced by model-\textsf{1}, and we can see that the reconstructed line by OHD is obviously excluded by this model. The rest three panels in Fig.\ref{fig:2} show the results by simulated future OHD with underlying models being $\Lambda$CDM, model-\textsf{1} and model-\textsf{2} respectively. Note that, for varying quality of data and underlying model complexity, $(N,M)$ are automatically adjusted, and are no longer suitable for current OHD, so we do not include current reconstructions. We can see that future OHD are able to reconstruct $w(z)$ very accurately, even for very oscillatory models.

\section{Discussion and conclusion}\label{sec.discussion}
We propose a new method by combining principal component analysis (PCA) and the goodness of fit (GoF) criterion to reconstruct dark energy equation of state (EoS) $w(z)$ by observational Hubble parameter data (OHD). We also used a new error model to simulate the error distributions of future OHD, and get forecasted simulated data by estimating potential surveys and data acquisition methods. In the GoF criterion analysis, we calculate the residual between current 28 OHD measurements and the concordance $\Lambda$CDM model and find that the residual ${\rm GoF}_\star\equiv\R_\star/n=0.62$ ($n=28$) is far from unity, which implies that these measurements should not be considered to be independent and thus improper to use a diagonal covariance matrix in the linear least squares. We bin the data by combining nearby redshift data and get newer 15 measurements of OHD and they seem to be independent -- their ${\rm GoF}_\star=1.02$. We note that this is only a rough estimation, and generally we need to get the covariance between the measurements. However, there is not a estimated covariance for OHD, so manually making them independent is just for comparing the power of current and future data on the EoS reconstruction.

We use our method to reconstruct EoS $w(z)$ by the current OHD, and discover a feature of deviation from $\Lambda$CDM at $z>1.5$. The Ia SN data give similar results \citep{2010PhRvL.104u1301C}. While the quality of current data cannot verify it, future data by our simulations greatly enhance the result (Fig.\ref{fig:2}), and are able to confirm or deny this deviation. In the analysis we used only cosmic chronometer data. We also use simulations with the data from other sources (BAO, SL signals), and the result does not improve much. Because the quantity of three BAO data and five SL data dominates only a small fraction of total 100 simulated data. However, with small number, say 20, of cosmic chronometer data, adding several SL signal data lowers the error of reconstruction at $z\simeq 2$, but beyond this redshift the result is still not confident. The reason is that, adding a few high-redshift data helps to determine $H'$ better at $z\simeq 2$ but is still unable to well determine $H'$ at $2\lesssim z\lesssim 5$. In such case we may use a derivative prior in Eq.(\ref{eq.beta}) or use a Gaussian process \citep{2012PhRvD..86h3001S} to help to determine $H'$.

\section*{Acknowledgements}
This work was supported by the National Science Foundation of China (Grants No. 11173006), the Ministry of Science and Technology National Basic Science program (project 973) under grant No. 2012CB821804, and the Fundamental Research Funds for the Central Universities.

\bibliographystyle{h-physrev3}
\bibliography{haoran_ref}

\end{document}